\documentclass[aps,prc,showkeys,reprint,superscriptaddress,onecolumn]{revtex4-2}

\usepackage{graphics}
\usepackage{graphicx}
\usepackage{xcolor}
\usepackage{lineno}
\usepackage{placeins} 
\usepackage[pdftex, pdftitle={Article}, pdfauthor={Author},colorlinks = true,allcolors = blue]{hyperref} 
\usepackage{comment}
\usepackage[colorinlistoftodos,prependcaption]{todonotes}
\begin{document}

\title{Grey Tracks as Probes of Hadronization Dynamics}


\author{Carolina M. {Robles Gajardo}}
\affiliation{Pontificia Universidad Cat\'olica de Valpara\'iso, Valpara\'iso, Chile}
\affiliation{Universidad T\'ecnica Federico Santa Mar\'ia, Valpara\'iso, Chile}
\affiliation{Centro Cient\'ifico y Tecnol\'ogico de Valparaíso, Valpara\'iso, Chile}
\affiliation{Instituto Milenio de F\'isica Subat\'omica en la Frontera de Altas Energ\'ias, Santiago, Chile}

\author{Alberto Accardi}
\affiliation{Hampton University, Hampton, VA 23668, USA}
\affiliation{Jefferson Lab, Newport News, VA 23606, USA}

\author{Mark D.~Baker}
\affiliation{Jefferson Lab, Newport News, VA 23606, USA}
\affiliation{MDB Physics and Detector Simulation LLC, Miller Place, NY 11764, USA}
\affiliation{Department of Physics, Brookhaven National Laboratory, Upton, New York 11973, USA}

\author{William K. Brooks}
\email{william.brooks@usm.cl}
\affiliation{Universidad T\'ecnica Federico Santa Mar\'ia, Valpara\'iso, Chile}
\affiliation{Centro Cientifico y Tecnol\'ogico de Valparaíso, Valpara\'iso, Chile}
\affiliation{Instituto Milenio de F\'isica Subat\'omica en la Frontera de Altas Energ\'ias, Santiago, Chile}
\affiliation{Department of Physics and Astronomy, University of New Hampshire, Durham, NH 03824, USA}

\author{Rapha\"el Dupr\'e}
\affiliation{Laboratoire de Physique des 2 Infinis Ir\`ene Joliot-Curie,
CNRS - Universit\'e Paris-Saclay, Orsay, France}

\author{Mathieu Ehrhart}
\affiliation{Laboratoire de Physique des 2 Infinis Ir\`ene Joliot-Curie,
CNRS - Universit\'e Paris-Saclay, Orsay, France}

\author{Jorge A. L\'opez}
\affiliation{Physikalisches Institut, Universit\"at Heidelberg, Heidelberg, Germany}

\author{Zhoudunming~Tu}
\affiliation{Department of Physics, Brookhaven National Laboratory, Upton, New York 11973, USA}
\date{\today}

\begin{abstract}
Energetic quarks liberated from hadrons in nuclear deep-inelastic scattering propagate through the nuclear medium, interacting with it via several processes. These include quark energy loss and nuclear interactions of forming hadrons. One manifestation of these interactions is the enhanced emission of low-energy charged particles, referred to as grey tracks. We use the theoretical components of the BeAGLE event generator to interpret grey track signatures of parton transport and hadron formation by comparing its predictions to E665 data. We extend the base version of BeAGLE by adding four different options for describing parton energy loss. The E665 data we used consists of multiplicity ratios for fixed-target scattering of 490 GeV muons on xenon normalized to deuterium as a function of the number of grey tracks. We compare multiplicity ratios for E665 grey tracks to the predictions of BeAGLE, varying the options and parameters to determine which physics phenomena can be identified by these data. We find that grey tracks are unaffected by modifications of the forward production. Thus their production must be dominated by interactions with hadrons in the backward region. This offers the advantage that selecting certain particles in the forward region is unlikely to bias a centrality selection. We see a strong correlation between the number of grey tracks and the in-medium path length. Our energy loss model does not reproduce the suppression observed in the projectile region. We see an underprediction of the proton production rate in backward kinematics, suggesting that a stronger source of interaction with the nuclear medium is needed for accurate modeling. These results lay an important foundation for future spectator tagging studies at both Jefferson Lab and at the Electron-Ion Collider, where neutron and proton grey track studies will be feasible down to very small momenta.
\end{abstract}

\maketitle
\section{Introduction}
The process by which energetic quarks and gluons propagate through space and evolve into hadrons, referred to as hadronization or fragmentation, has not yet been understood in terms of the Lagrangian of Quantum Chromodynamics (QCD). The first ingredients for quantitative study of the process are the parton distribution functions (PDFs). Although PDFs for the proton have been studied for decades, new insights into their estimation continue to be unearthed~\cite{BRODSKY2022136801}. For scattering from small objects such as the proton, the formalism of QCD factorization~\cite{Collins:1989gx} in such reactions as semi-inclusive deep inelastic scattering (SIDIS) allows the observed final state to be parameterized as the convolution of PDFs with the conventional fragmentation functions~\cite{Hirai:2007cx} within many kinematic conditions~\cite{Pasechnik:2011nw}, as successfully demonstrated by a recent global extraction of fragmentation functions from many different experiments~\cite{Borsa:2021ran}. However, the information provided on hadronization in that scenario is limited to quantities such as hadron multiplicities and particle production ratios, such as $\mathrm{K}/\pi$ ratios.

Access to completely new information on hadronization can be obtained by implanting the process inside atomic nuclei \cite{HIJINGPhysRevD.44.3501}. While the long distance nature of the interactions within the nucleus might be presumed to preclude naive application of factorization-based methods\cite{BRODSKY2022136812}, there is nonetheless recent progress in extending the usual approaches for proton targets to describe data for parton distribution functions in nuclei (nCTEQ15 \cite{nCTEQ15_PhysRevD.93.085037}, EPPS16 \cite{EPPS16_2017}, nNNPDF2.0 \cite{nNNPDF2_2020}, and TUJU19 \cite{TUJU19_PhysRevD.100.096015,ABMP16_2018}) even at low energies and high-$x_{\text{Bj}}$ \cite{Paukkunen:2020rnb,Segarra_PhysRevD.103.114015}.

In this paper we press forward on a new and promising front to understanding hadronization in cold nuclear matter, namely the study of so-called grey tracks, protons in the range momentum 0.2--0.6 GeV$/c$, in lepto-nuclear scattering. This historical term comes from tracking technologies such as nuclear emulsion or streamer chamber measurements, where the visual appearance of the measured charged particle track differs from that of the highest energy particles. Grey tracks ($n_{\text{g}}$) correspond to lower momentum charged particles with an ionization energy loss much greater than that of minimum ionizing particles. In the context of this paper, we are studying them as a proxy for the interactions of the energetic parton and/or forming hadron with the nuclear medium. The factorization theorem states that the hadronization of a quark or gluon with high momentum is independent of the target, projectile and subprocess \cite{Collins:1989gx}. 
By performing detailed modeling of grey track production  with the BeAGLE event generator enhanced by the upgrade of the PyQM module, we can frame the problem of getting microscopic information on such processes from lepto-nuclear scattering such as the data from the E665 fixed-target experiment at Fermilab. This work lays the foundation for future studies at fixed-target experiments at Jefferson Lab and other electron-beam facilities, but more importantly, at future lepton collider experiments such as the proposed electron-ion colliders \cite{Accardi:2012qut,Anderle_2021,FCC:2018bvk,FCC:2018byv,FCC:2018evy,FCC:2018vvp}. In the context of collider experiments, it is feasible to study extremely low momentum particles such as grey tracks because they emerge from the nucleus having a velocity in the lab frame very close to that of the hadron beam. The charged component can be momentum analyzed using the magnetic fields already present for the recirculating beam transport, while the neutral component can be measured using the zero-degree calorimeter technique \cite{Leite:2013qxa,Suranyi:2021ssd,khalek2021science}. These techniques, sometimes referred to as spectator tagging or geometry tagging \cite{Morozov:2018voq,Zheng:2014cha,ExposingNovel}, often require special instrumentation to detect very low-energy particles associated with highly energetic interactions\cite{Dupre:2017upj,FENKER2008273}. Geometry tagging studies of grey tracks provide unprecedented new information on the microscopic interactions of the hadronization constituents with the cold nuclear medium. 

The concept for studying grey tracks in experimental measurements originated in the 1960's~\cite{Feinberg1966} and was discussed through the following decades~\cite{Dar:PhysRevD.6.2412} generally in the context of hadron-nucleus collisions, although it was recognized that deep inelastic scattering by lepton beams would be simpler to interpret~\cite{Goldhaber:PhysRevD.7.765}. An experimental exploration of these ideas took place for proton-nucleus interactions in the E910 experiment at the Alternating Gradient Synchrotron at Brookhaven National Lab beginning in 1996, measuring slow protons and deuterons produced in the collison of 18 GeV/c protons with Be, Cu, and Au targets~\cite{PhysRevC.60.024902}. In comparing their data with several models, they concluded that there was a strong linear dependence of the number of projectile-nucleus  interactions $N_{\text{pA}}^{\text{int}}$ on the mean number of grey tracks measured, with a target-dependent constant of proportionality. Defining $N_{\text{pA}}^{\text{int}}$ as the centrality of the collision, they concluded that measurement of the mean number of grey tracks can determine event centrality well for a given nucleus.

The E665 experimental collaboration published more than 20 papers in the 1990's based on measurements with a secondary muon beam of 490 GeV at Fermi National Accelerator Laboratory. One of these included measurements of grey tracks produced from a xenon target, making a comparison to those seen from a deuteron target \cite{Adams:1995}. These data stimulated renewed theoretical interest in using grey tracks to study the space-time development of hadronization via the nuclear medium. References~\cite{CIOFIDEGLIATTI2005281} and \cite{CiofidegliAtti:2005mp} explored the dependence of the mean number of grey tracks on four-momentum transfer $Q^2$ and $x_{\text{Bj}}$ within a theoretical model. They found that the number of grey tracks rises with increasing $Q^2$, and found fair agreement with the dependence measured in E665.

\section{The E665 Data}
The E665 data being described by the simulations in this paper consisted of muon-xenon scattering compared to muon-deuterium scattering in Deep Inelastic Scattering (DIS) kinematics. The experiment achieved nearly 4$\pi$ steradians of acceptance for charged particles by using a streamer chamber immersed in a 15 kG dipole magnetic field as a vertex detector. This enabled study of grey tracks in combination with the high-energy forward tracks, which were analyzed with a downstream spectrometer system. Scattered muons were identified with a downstream iron absorber followed by tracking devices. An electromagnetic calorimeter was used for photon detection to identify muon bremsstrahlung. 

The experimental definition of grey tracks used by E665 starts with identification of the scattered muon in the downstream muon detector, with track matching to the trajectories measured upstream. Charged particles not identified as muons were treated as hadrons. Subsequently, a rigorous visual analysis of the photographs of hadron tracks in the streamer chamber was performed. As low momentum particles, they deposited much more energy per unit length than the minimum ionizing particles, thus leaving a much higher streamer density. Grey tracks were counted independently in two laboratories, and these analyses agreed to within 11\%. In addition to this visual analysis, they were required to have  momentum in the interval 0.2 - 0.6 GeV/c. The majority of these particles were protons.

\section{Geometry Tagging for DIS }
The term ``Geometry Tagging" in this context refers to the measurement of particles emerging from the nucleus to determine geometric characteristics of the hard interaction and of the subsequent final-state interactions of the participants, including spectator particles such as evaporation neutrons, fission fragments, hadrons produced from gluonic bremsstrahlung from in-medium partons, and low energy knocked-out protons and neutrons, among others. BeAGLE is able to simulate such processes, and it has been employed in evaluations of the effectiveness of determining geometric features from various experimental signatures, such as those just mentioned, for collider experiments~\cite{Morozov:2018voq}. That work follows earlier studies with the DPMJET3 code for forward neutron tagging~\cite{Zheng:2014cha} in which it was found that the evaporation neutrons could give information on the event centrality.

The impact parameter \textbf{b} is a well-known feature of the scattering process. It is particularly relevant for hadron-nucleus scattering since the probability of interaction of the incident hadron with the medium is large due to the magnitudes of hadronic cross sections. Thus, the projectile pathlength through the medium is strongly correlated to the impact parameter, and the number of interactions of the projectile within the medium is indicative of the centrality of the collision, as studied by the E910 collaboration discussed earlier. For lepton beams, which have electroweak cross sections, a more relevant quantity is the pathlength in the medium of the energetic quark and any subsequent hadron that contains it. This is because for higher $x_{\text{Bj}}$ interactions, leptons and photons easily penetrate into the nucleus so that the hard interaction can take place anywhere within its volume. For the energetic quark, the quark pathlength, labeled \textbf{d}, can be small for multiple reasons. It can be small because its color lifetime (the time interval between the  first interaction of the quark with the medium and the neutralization of its color) is short, reflecting that its most probable value is zero as is the case with ordinary lifetimes, or it can be because the interaction is on the periphery of the nucleus. However, it can only be longest for central collisions for events in which the color lifetime has a large value. The color lifetime can be large either because it fluctuates to a large value, or because of kinematic conditions such as a relative energy $z_\mathrm{h} \equiv E_\mathrm{h}/\nu$ that is near the optimal value of $\approx 0.3$, as indicated by the Lund String Model prediction~\cite{Brooks:2020fmf}. As will be discussed later, in BeAGLE we find that on average a larger value of \textbf{d} is correlated with a larger value of the number of grey tracks. A maximal value of \textbf{d} is more suggestive of a colored energetic quark with a long color lifetime in a central collision that produces low energy particles through medium-induced energy loss, and less consistent with an inelastic hadronic collision, which would be more likely to produce higher energy hadrons. 

\begin{figure*}[t!] 
    \centering
    \includegraphics[width=9cm]{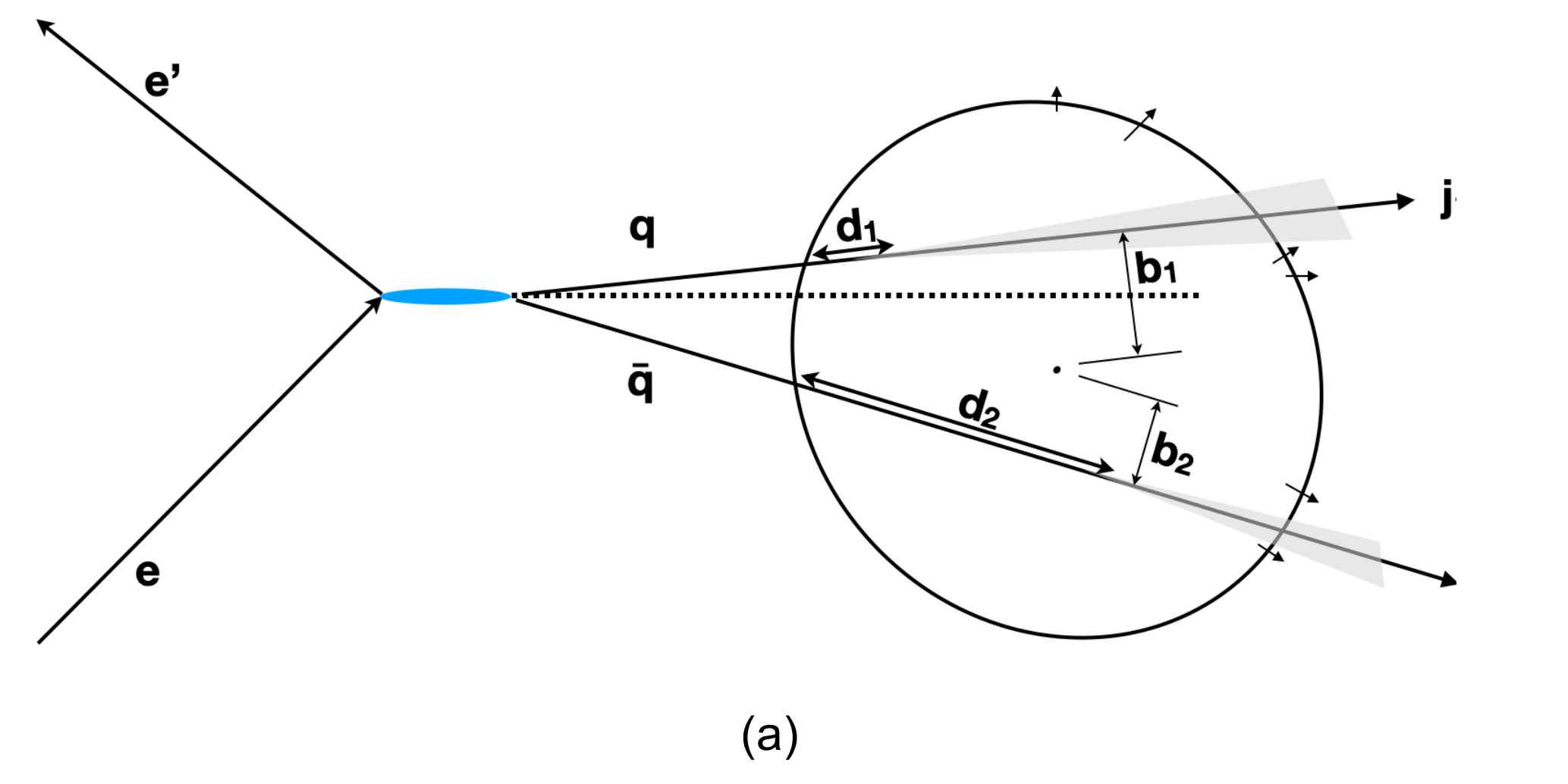}
    \includegraphics[width=7cm]{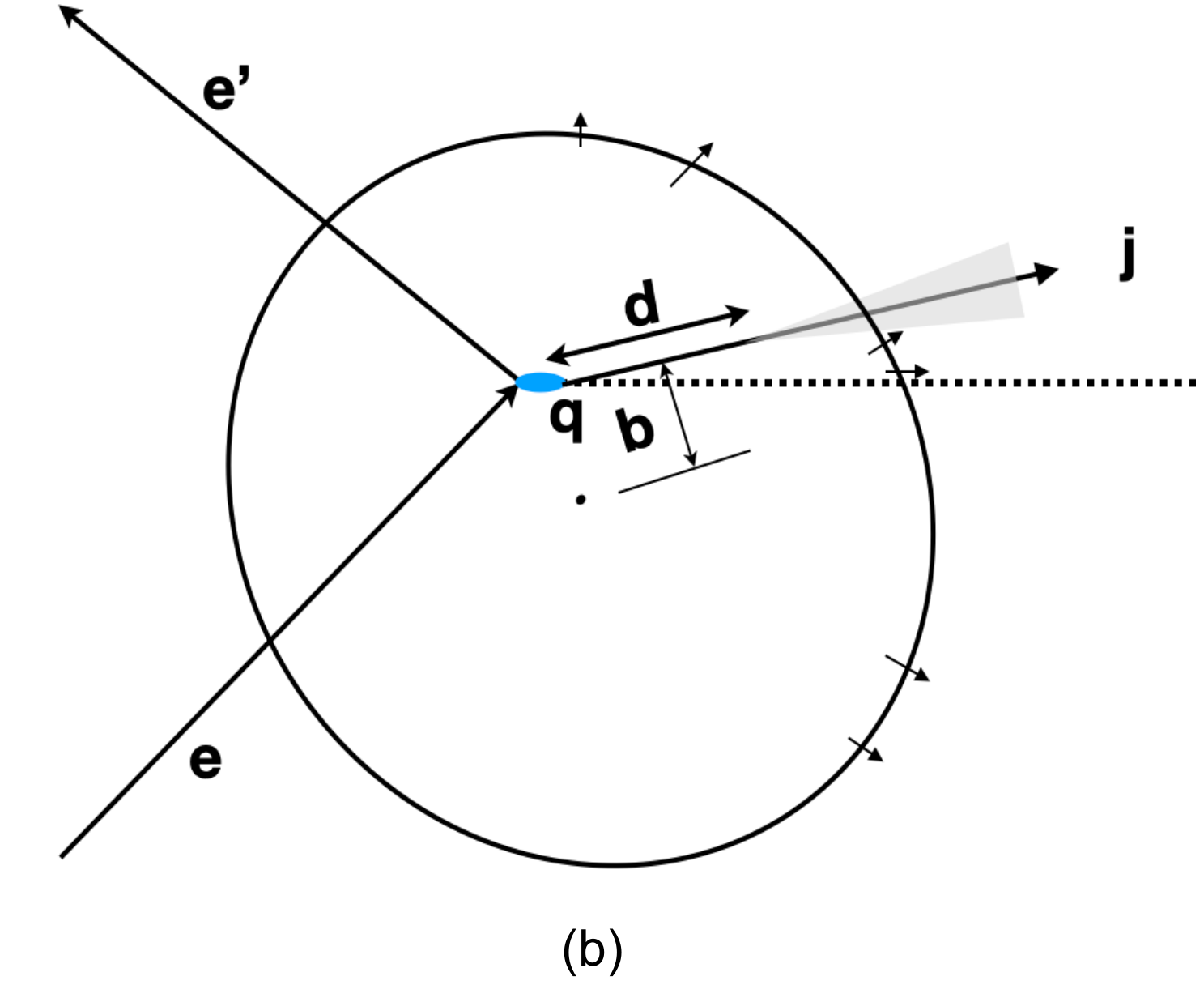}
    \caption{Diagrams illustrating geometric features of the leptonuclear scattering process in the Target Reference Frame. As depicted in (a), for this particular event the quark $q$ with impact parameter $b_1$ has a lower energy which on average results in a shorter color lifetime $d_1/c$, and the antiquark $\bar{q}$ with impact parameter $b_2$ has a higher energy which on average results in a longer color lifetime $d_2/c$. The propagating quarks begin the evolution into jets or hadrons $j_1$ and $j_2$ either inside or outside the medium. In (b) is shown the simpler process that dominates at higher $x_{\text{Bj}}$. The energy and momentum of the virtual photon is absorbed by a single valence quark $q$ with impact parameter $b$ which evolves into jet or hadron $j$. The degree of misalignment between the virtual photon direction and the direction of leading hadron or jet \textbf{j} due to intrinsic $k_\perp$ and Fermi momentum is  exaggerated in the right-hand diagram (b) to facilitate visualization. In (b) it is possible to speak of the ``struck quark'' $\gamma^*+q \to q$ while (a) corresponds to $\gamma^*+g \to q\bar{q}$, see text. Small arrows in both diagrams represent low-energy hadrons emitted from the nucleus due to various processes, e.g., potential grey tracks. Features not shown in the figure include: the falloff of the nuclear density at the boundary of the nucleus; other quarks and gluons produced in the process; possible fission of the nucleus; and neutron evaporation from the nuclear fission fragments.}
    \label{fig:geometry_diagram}
\end{figure*}
In Fig. \ref{fig:geometry_diagram} is shown an illustration of \textbf{b} and \textbf{d} for low $x_{\text{Bj}}$ and for high $x_{\text{Bj}}$, adapting the development of references~\cite{DelDuca:1992ru,BrodskyHoyerPrivComm}. In what follows we primarily consider the color propagation in the Target Rest Frame (TRF), while clarifying the  connections to the Infinite Momentum Frame (IMF) for completeness. At lower $x_{\text{Bj}}<0.1$, which is the case for most of the E665 data, the quark pair production process becomes possible. The process of producing a virtual photon and $q \bar{q}$ pair in the target reference frame is distributed over a longer distance in space due to the Ioffe time $\approx 1/(2\cdot x_{\text{Bj}}\cdot{M_\mathrm{p}})$ or coherence length~\cite{Kovchegov:2001dh}, with the pair formation naturally occurring as one of the DGLAP radiation loop processes that are proportional to $\mathrm{log}(Q^2)$. For higher $x_{\text{Bj}}>0.1$, the simpler process of absorption of the virtual photon by a valence quark, shown on the righthand side of the figure, is dominant.

In the TRF the DIS cross section $\sigma_{\text{DIS}}$ is determined by the scattering of the $q \bar{q}$ Fock state of the virtual photon on the target (and of higher Fock states at higher orders of $\alpha_s$). At leading twist there are two qualitatively distinct configurations of the $q \bar{q}$ pair which give rise to $\sigma_{\text{DIS}} \propto 1/Q^2$:

\begin{enumerate}
    \item {\textbf{Quarks with similar momenta:}}
    the two quarks share the photon longitudinal momentum $\sim \nu$ roughly equally, and have relative transverse momentum $\sim Q$. Then the transverse size of the pair is $\sim 1/Q$, and their scattering cross section in the target is $\propto 1/Q^2$ (color transparency). The pair dominantly scatters through gluon exchange to the target. In the standard IMF frame this corresponds to the physical process $\gamma^*+g \to q \bar{q}$, which occurs via higher orders of the photon's Fock space. The gluon needs to be highly virtual to resolve the small-sized quark pair, hence this process is suppressed by $\alpha_s(Q^2)$. Nevertheless this process dominates at low $x_{\text{Bj}}$ due to the large gluon distribution, corresponding to the left-hand side of Fig. 1.
    \item{\textbf{Quarks with very different momenta:}}
    one of the quarks takes nearly all of the photon momentum $\nu$, while the other’s momentum remains fixed in the Bjorken limit. The transverse size of such pairs originates in the intrinsic $k_{\perp}$ distribution of the parton, $\sim 1$~fm, and so they scatter with a large ($Q$-independent) cross section. The probability of such a Fock state in the photon is $\propto 1/Q^2$, accounting for  $\sigma_{\text{DIS}} \propto 1/Q^2$. The quark with fixed momentum may be considered to be part of the target wave function. Thus this process corresponds, in the IMF, to the lowest order parton model $\gamma^*+q \to q$ shown on the right-hand side of Figure 1. 
\end{enumerate}

There is a smooth connection between cases 1 and 2. At low $x_{\text{Bj}}$ the quark with $Q$-independent momentum nevertheless has a large longitudinal momentum $\propto 1/x_{\text{Bj}}$, and starts to interact via gluon exchange. Such quarks correspond to sea quarks in the IMF, generated by gluon splitting at higher orders of $\alpha_s$.

In the following section we will make use of the PYTHIA6 module within BeAGLE to quantify the relative contributions of these processes in the E665 kinematics.

\section{Monte Carlo simulation: BeAGLE}
The Monte Carlo code ``Benchmark eA Generator for LEptoproduction" (BeAGLE)~\cite{Chang:2022hkt} version 1.02 was used to simulate the E665 grey tracks in this work. BeAGLE is a general purpose FORTRAN program for simulating electron-nucleus (eA) interactions. We are testing and using the PyQM module which we updated in the current version of BeAGLE to explore the sensitivity of the grey track observables to partonic energy loss in the cold nuclear medium. 

BeAGLE is a hybrid model that uses the DPMJet~\cite{DPMJET:2000}, PYTHIA6~\cite{PYTHIA6:2006}, PyQM, FLUKA~\cite{10.3389/fphy.2021.788253,BATTISTONI201510}, and LHAPDF5~\cite{Whalley:2005nh} codes to describe high-energy leptonuclear scattering. The geometric density distribution is provided primarily by PyQM while the quark distributions within that geometry are provided by nPDF EPS09~\cite{EPS09:2009}. The parton-level interactions and subsequent fragmentation is carried out by PYTHIA6. Hadronic formation and interaction with the nucleus is described by DMPJet, the impact of the scattering on the nucleus is described by FLUKA, including nucleon and light fragment evaporation, de-excitation by photon emission, nuclear fission, and Fermi breakup of the decay fragments. The PyQM module implements the Salgado-Wiedemann quenching weights to describe partonic energy loss~\cite{SW:2003}.

BeAGLE includes a variety of options to control the phenomena included in the simulation, some of which are mentioned here. Nuclear shadowing is described via two different approaches. The hadron formation time is accounted for in the DPMJet intra-nuclear cascade. Fermi motion of the nucleons in the nucleus can be described with several different mechanisms, or turned off completely. The nuclear geometry parameters from PyQM can be overridden, and deformed nuclei can be described. The PyQM actions available include specification of the $\hat{q}$ transport coefficient to adjust the degree of interaction between energetic partons with the nuclear environment. Some details of the partonic energy loss process in PyQM can also be selected, such as the fraction of the recoil in PyQM transferred to the nucleus, the modeling options for quark/hadron  transverse momentum generated by the medium, the multiplicity and energy of the generated gluons, and other options.
The main program is DPMJet, which uses PYTHIA6 to handle elementary interactions and fragmentation. PyQM handles this directly after elementary interactions in PYTHIA6, while DPMJet handles nuclear geometry and, after fragmentation by PYTHIA6, DPMJet takes care of the nuclear evaporation by FLUKA.


As described earlier, at high $x_{\text{Bj}}$ the process $\gamma^*+q \to q$ dominates, and at low $x_{\text{Bj}}$ the process $\gamma^*+g \to q \bar{q}$ dominates. Here we quantify the relative probability of those two kinds of processes according to the modeling of PYTHIA6 in the E665 kinematics. These are quantified in three categories, following Reference \cite{Adams:1995}: $x_{\text{Bj}}<$ 0.02 (shadowing region), $x_{\text{Bj}}>$ 0.02 (no shadowing region), and the full range accessed 0.002 $<x_{\text{Bj}}<$ 0.3.
  
\begin{table*}
   \centering
   \caption{Percentage of the processes in which a ``point photon'' is used and in which  interaction mechanisms are used in PYTHIA6 for the BeAGLE parameter settings used in this work. The BeAGLE hadron formation time parameter tau0 was set to 7 fm. For the ``point photon'' processes, the hard-scattering final state is composed of a single $q$ or $\bar q$ (LO), a $q\bar{q}g$ triplet (QCD Compton), or a $q\bar{q}$ pair (PGF). In all cases an appropriate soft nucleon remnant or remnant cluster is produced by PYTHIA6. For ``vector meson production'' the final state is $V+N,V*+N,V+N*,V*+N*$, where $*$ means a diffractively broken up state. In the ``resolved process,'' the $\gamma^*$ is treated as a hadron and interacts through a hard QCD process with the nucleon. Thus, a quark, antiquark or gluon from the $\gamma^*$ interacts with a quark, antiquark or gluon from the nucleon, leaving two remnant systems. The ``low-$p_{\text{T}}$'' processes collect together remaining  categories that contribute to low $p_{\text{T}}$ interactions. The numbers in this table are calculated for xenon, but they are very similar for deuterium.}
 
  \begin{tabular*}{1.0\textwidth}{@{\extracolsep{\fill}}lcccc@{}}
    \hline
     Process & $x_{\text{Bj}}<$ 0.02 & $x_{\text{Bj}}>$ 0.02 &  0.002 $<x_{\text{Bj}}<$ 0.3 &   \\ 
      \hline
       Point-photon: LO-DIS & $85.58\%$ & $98.37\%$ & $93.46\%$ & \  \\
        Point-photon: QCD Compton & $3.98\%$ & $0.65\%$ & $1.92\%$ & \  \\
         Point-photon: Photon-Gluon Fusion & $7.42\%$ & $0.41\%$ & $3.09\%$ & \  \\
         Vector meson production & $0.41\%$ & $0.13\%$ & $0.24\%$ & \  \\
     Resolved processes & $2.17\%$ & $0.16\%$ & $0.93\%$ & \  \\
    Low $p_{\text{T}}$ processes & $0.44\%$ & $0.28\%$ & $0.34\%$ & \  \\
  
  \end{tabular*}
   \label{tab:BigSummarySystematics}
\end{table*}

As is evident in Table 1, the PYTHIA6 assessment of the various processes indicates  that the high-$x_{\text{Bj}}$ process described in the previous section, and shown in Fig.~\ref{fig:geometry_diagram} (right), is the dominant process by far. Even though the E665 data go as low as 0.002 in $x_{\text{Bj}}$, according to PYTHIA6 it is not low enough to have a significant production of the low-$x_{\text{Bj}}$ process described in the previous section and shown in Fig.~\ref{fig:geometry_diagram} (left). This will presumably be much more prevalent in high energy colliders such as the future high energy Electron-Ion Collider.

\section{Medium-induced gluon radiation: PyQM}
In nuclear DIS processes, the partons emerging from hard scattering will, for a time $\tau_{\text{c}}$, travel as a colored particle \cite{Accardi:2009zzb}, and experience multiple rescatterings on the nucleus remnant. The parton energy degradation -- or, parton energy loss -- caused by medium-stimulated gluon radiation is an important process to understand the suppression of large momentum hadron production observed in semi-inclusive nuclear DIS processes at various facilities~\cite{Accardi:2009qv}.

In BeAGLE, the PyQM module is used to simulate medium-induced gluon radiation and quark energy loss. For each quark and gluon generated by PYTHIA6, the amount $\Delta E$ of energy radiated by a parton traveling through a nucleus is stochastically generated using ``quenching weights'' calculated by Salgado and Wiedemann~\cite{SW:2003}, that provide the probability distribution $P(\Delta E;\omega_{\text{c}},R)$ in the energy $\Delta E$ . The model considers a static and uniform medium of length $L$ with scattering power quantified by the transport coefficient $\hat{q}$, that measures the average parton transverse momentum broadening per unit path length. The radiation process is regulated by the gluon characteristic energy $\omega_{c}=\frac{1}{2}\hat{q}L^{2}$. In a finite size medium, large angle radiation suppression is controlled by the cutoff parameter $R=\omega_{\text{c}} L$. The geometry of the nuclear medium is taken into consideration by using a Wood-Saxon density distribution~\cite{DeVries:1987atn} to randomly generate the struck quark production point, and calculate its average in-medium path length $d$. The energy loss $\Delta E$ is then calculated using quenching weights with $L=d$. The partonic final state generated by PYTHIA6 is then modified by removing the appropriate amount of energy from each of the partons generated in the hard scattering, and furthermore adding gluons to the final state, or letting these be absorbed by the nucleus remnant as discussed below. Finally, hadronization of this modified final state is handled as usual by PYTHIA6's implementation of the Lund string fragmentation model. A more detailed description of PyQM can be found in~\cite{Dupre:2011afa}. The in-medium interactions of non partonic states generated by PYTHIA6 -- especially soft hadrons -- are handled by BeAGLE's DPMJET module. 

We have implemented four different energy loss scenarios when interfacing PyQM to the BeAGLE simulation. These differ in the way the radiated gluons are added to the final state generated by JETSET and how much of their energy is absorbed by the nuclear medium, or escapes it.

In the first option, called \textbf{no gluons}, there is no compensation for the energy lost by the hard partons, namely, we neither add gluons to the final system to carry away the lost energy, nor do we redistribute this in the rest of the nuclear remnant. Energy conservation is therefore broken, and we use this option for cross-check purposes. 

In the second option, called \textbf{1-hard gluon}, we attribute the energy loss to the radiation of a gluon generated in a single hard re-scattering on the nuclear medium. This gluon is added to the final state with energy  $\Delta E = E_{\text{parton}}^{\text{initial}}-E_{\text{parton}}^{\text{final}}$, where $E_{\text{parton}}^{\text{initial}}$ is the initial energy of the parton that will lose energy, and $E_{\text{parton}}^{\text{final}}$ is the result of its energy loss, isotropically distributed as transverse momentum of magnitude $q_{\text{T}}^2 = \hat{q}L$, and a longitudinal momentum appropriate for an on-shell massless gluon. 

In the third case, called \textbf{1-hard + soft gluons}, we consider the energy of a radiated hard gluon based on the multiple soft scattering approximation \cite{SW:2003}, where the gluon energy spectrum per unit path length is:
\begin{equation}
\label{eq:sw2003}
    \omega\frac{d I}{d\omega dz}\simeq \alpha_{s}\sqrt{\frac{\hat{q}}{\omega}} 
\end{equation}
for $\omega < \omega_{\text{C}}$. After integrating over the path length and $\omega$, along with considering that the multiplicity of gluons emitted with energies larger than $\omega$ is $N(w')=\int_{w'}^{\infty }dw'\frac{d I(w')}{dw'}$ we obtain for one gluon:
\begin{equation}
\label{eq:whard}
    \omega_{\text{hard}}=4\alpha_{s}^{2}L^{2}\hat{q}
\end{equation}
where $L$ is the average path length, $\hat{q}$ is the transport coefficient and $\alpha_{s}$ is the strong coupling at soft scales. The energy loss is calculated as $\Delta E = E_{\text{parton}}^{\text{initial}}-E_{\text{parton}}^{\text{final}}$, just like the option above, which gives the total energy available to us as $\Delta E$, where $\omega_{\text{hard}}$ can be either less than or equal to $\Delta E$. To convert energy in cases where $\omega_{\text{hard}}$ is less than $\Delta E$, we create soft gluons with energy:
\begin{equation}
\label{eq:wsoft}
    \omega_{\text{soft}}=\Delta E - \omega_{\text{hard}},
\end{equation}
For technical reasons, FLUKA has difficulties to handle very high energy deposited in the nucleus. To circumvent this issue, we add a constraint on the soft gluons that they must not have energy greater than 5 GeV. In the scenario where they can have this energy or more, we produce a $qg\bar{q}$ triplet with energy:
\begin{equation}
\label{eq:wtriplet}
    \omega_{\text{triplet}}=\omega_{\text{soft}} - 5~[\text{GeV}].
\end{equation}
In summary, in this option we add a hard gluon with energy $\omega_{\text{hard}}$ while the remaining radiated energy is transferred to the remnant nucleus and a $qg\bar{q}$ triplet can be created in case of energy excess. 


In the last option, called \textbf{soft-gluons}, we do not add a gluon to the PYTHIA6 list, but we conserve energy by redistributing the energy loss to the remaining nuclei. We model the process assuming that only soft gluons are radiated with a total energy $\Delta E = E_{\text{parton}}^{\text{initial}}-E_{\text{parton}}^{\text{final}}$ which is redistributed to the nuclear remnant. Similarly to the previous option, we have to limit the total energy sent to the nucleus and will create a triplet to absorb any extra energy beyond 5~GeV. This option is similar to option one, with the only difference that we conserve the energy.

As an additional remark, for the purposes of calculation of energy loss, we assume that the photon energy, proportional to $\nu$, is large enough to boost the struck parton's lifetime outside of the nuclear target. Consequently, $\tau_c \gg d$, and induced gluon radiation occurs all along the struck parton's path through the whole nucleus. This should be a fair approximation for forward tracks at the energy of the E665 experiment considered in this work, as well as at the future Electron-Ion Collider. It is not necessarily true for the target projectile region, even in E665 kinematics, since there is strong experimental evidence of intranuclear cascades associated with backward kinematics. BeAGLE uses an estimate for $\tau_c$ to decide when each hadron is formed, and if hadrons are produced inside the nucleus, a detailed simulation of the resulting intra-nuclear cascade is initiated. At lower energies, such as at Jefferson Lab, hadronization may occur inside the medium even for forward tracks, and one would also need to include prehadron formation and nuclear absorption in the simulation.

 \begin{figure*}[t!]
    \centering
    \includegraphics[width=\linewidth,angle=0]{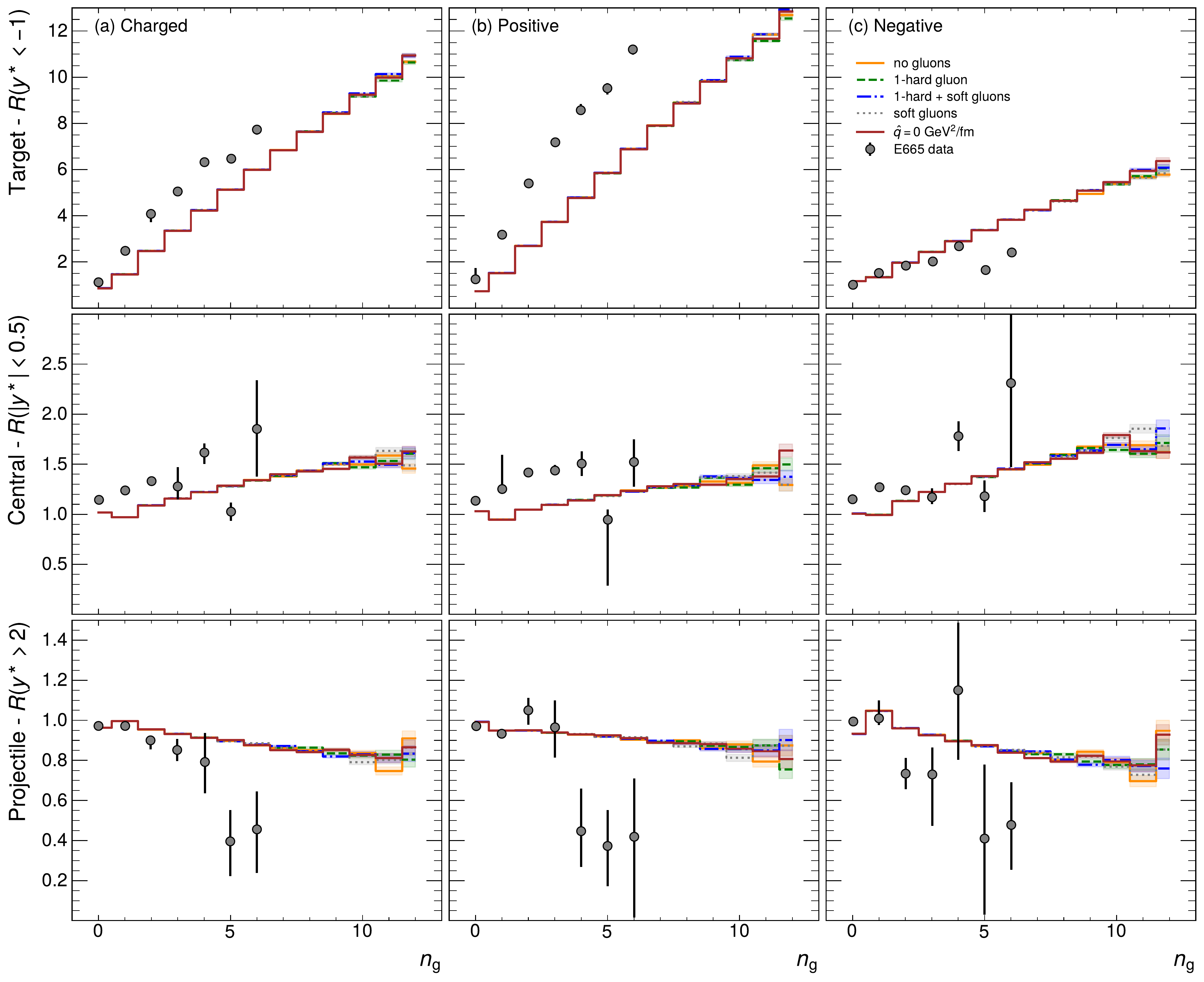}
    \caption{Multiplicity ratio $R(n_{g})_{\mu \text{Xe}}$ as a function of the number of grey tracks $n_{\text{g}}$ for charged, positive and  negative hadrons, in three rapidity intervals: the target, central and projectile fragmentation regions, and for four PyQM options with transport coefficient $\hat q = 0.5$~GeV$^2$/fm. The different columns correspond to charged, positive and negative hadrons and they are drawn in (a), (b) and (c), respectively. The different options of PyQM are highlighted by different types of lines and colours, where the orange (upper) solid line corresponds to the no gluons option, the green dashed-dotted line corresponds to 1 hard gluon option, the blue dashed-dotted line corresponds to 1 hard + soft gluons option, grey dotted line corresponds to soft gluons options, red (bottom) solid line corresponds to the simulation without induced energy loss with $\hat q = 0$~GeV$^2$/fm and full grey circles correspond to the E665 data.}
    \label{fig:multiplicity-ratio-1}
\end{figure*}
\section{Results: Comparison to E665 experimental data}
The analysis \cite{Adams:1995} made by the E665 experiment compares $\mu \text{Xe}$ and $\mu \text{D}$ data and $p \text{Xe}$ scattering from the NA5 collaboration to study nuclear effects in the hadronic system, with the grey tracks in an event being the main tool in the comparisons. They reported that the average multiplicity $\langle n_{\text{g}} \rangle$ of the grey tracks is significantly lower with $\mu \text{Xe}$ scattering than with $p \text{Xe}$ scattering, consistent with expectations due to the absence of initial state interactions between the incident particle and the nucleus in the muon scattering data.
 
 The E665 measurements were performed with a muon beam with an average energy of 490~GeV. Since particle identification detector information was not available, a partial identification of protons was used based on the physics expectation, considering that in $\mu \text{D}$ scattering with $E_{\mu}=490$~GeV and $W>8$~GeV about $50\%$ of the positive hadrons with $x_{\text{F}}(m_{\pi})<-0.2$ were protons \cite{Adams:1995}.
 
 To define the DIS region, to avoid kinematic regions where the radiative corrections are large, and to exclude the region where the experimental resolution is poor, they applied the following kinematic cuts: $Q^{2}>1$~GeV$^{2}$, $8 < W < 30$~GeV and $x_{\text{Bj}}>0.002$. After all cuts, the total number of events in BeAGLE is approximately 4.6 million events for deuterium target and 4.5 million events for xenon target, and 6309 events for deuterium target and 2064 events for the xenon target in Adams et al.\cite{Adams:1995}. Although in the data the grey tracks were assumed to be predominantly protons, there was a large contamination by pions and kaons, estimated at about $40\%$ and $(15 \pm 9)\%$ for the $\mu \text{D}$ and $\mu$Xe data, respectively\cite{Adams:1995}.
 
 In the above reference, a comparison was made between the results for multiplicity ratios with and without grey tracks. It was concluded that grey tracks are very efficient for tagging events where cascade interactions occur because the nuclear effects on hadrons are strongly enhanced in the sample of $\mu \text{Xe}$ events containing grey tracks. The sample of $\mu \text{Xe}$ events without grey tracks appeared very similar to the sample of $\mu \text{D}$ events, reinforcing the idea that diffractive scattering has a very similar pattern on deuterium and xenon.

 In BeAGLE, the same kinematics cuts were applied, with the only exception that the grey tracks $(n_{\text{g}})$ are directly identified as protons with momentum $0.2 < p < 0.6$~GeV.
 
 The multiplicity ratio~\cite{Adams:1995} was defined as 
 \begin{equation}
 \label{eq:adams1995}
    R=\frac{\langle n(n_{\text{g}}) \rangle_{\mu \text{Xe}}}{\langle n \rangle_{\mu \text{D}}}
 \end{equation}
where $\langle n(n_{\text{g}})\rangle _{\mu \text{Xe}}$ is the average number of hadrons in the final state including charged pions and kaons, protons, antiprotons   in the xenon target, and $\langle n\rangle _{\mu \text{D}}$ is the average number of hadrons and leptons in the final state for the deuterium target. Moreover, the same charge cuts were applied to the denominator as to the numerator, and grey tracks were also included in $\langle n(n_{\text{g}})\rangle _{\mu \text{Xe}}$ and $\langle n\rangle _{\mu \text{D}}$.

E665 concluded that there were significant diffractive scattering contributions to these data. They estimated that a lower bound of 18 $\pm$3\% of all events were diffractive for $x_{\text{Bj}}$ less than 0.02 on xenon. It can be anticipated that diffractive events would modify the measured multiplicity ratios. It is also commented in section 4.8 of \cite{Adams:1995} that diffractive scattering events would have a similar topology for xenon as for deuterium. It goes on to say that such scattering events, which involve particle combinations with the quantum numbers of the vacuum, would naturally produce fewer hadrons than the DIS processes, which involve color octet particles in propagation, on average. As a result, the multiplicity ratio, defined in eq. 5, changes its value if diffractive scattering occurs. The numerator and denominator of the multiplicity ratio are both affected by the presence or absence of diffractive scattering. The numerator, an average over the multiplicity for xenon as a function of the number of grey tracks, would be enhanced by diffractive scattering for low n$_g$. For example, it contributes more events to the n$_g$=0 bin than the pure DIS process would, on average. In the denominator, which is the average deuterium multiplicity, the diffractive scattering would also produce cleaner events than DIS, biasing the average to smaller numbers. If this trend were to be corrected, the denominator should be made larger. 

Therefore, the expected effect on the multiplicity ratio, relative to pure DIS events, would be that the numerator is too large for low-n$_g$ events, and the denominator is too small for all measured events. Both of these effects imply that for  pure DIS events, the multiplicity ratio should be smaller, particularly for low-n$_g$ events, with the n$_g$=0 bin being the most strongly affected. As can be seen in Fig. 2, a somewhat smaller multiplicity ratio due to these effects could improve the overall agreement of BeAGLE with the E665 data. We do not attempt to make a correction to account for the effects due to diffractive scattering; the trigger used for acquiring these data, which required multiple charged particles, tended to partly suppress coherent diffractive scattering to an extent that is difficult to estimate.

The regions are divided into different values of rapidity calculated in the photon-lepton center of mass frame as 
 \begin{equation}
     y^{*}= \ln \left( \frac{E_{\text{h}}+p_{\text{L}}^{\text{h}}}{E_{\text{h}}-p_{\text{L}}^{h}} \right),
 \end{equation}
 where the first row is the target fragmentation region with rapidity values $y<1$, the middle row is the central fragmentation region with rapidity values between $-0.5<y<0.5$ and the last row is the projectile fragmentation region with rapidity values $y>2$. Each column indicates the charge of the hadrons and leptons in their final state: the first column stands for charged hadrons and leptons, the second for positive hadrons and the third for negative hadrons and leptons.

The E665 collaboration observed that the multiplicity ratio $R$, eq.~\ref{eq:adams1995}, with $n_{\text{g}}$ in different rapidity regions, reveals diverse characteristics of particle production on a nucleus. In Fig.~\ref{fig:multiplicity-ratio-1}, the multiplicity ratio as a function of the number of grey tracks $n_{\text{g}}$ is shown. The grey circles represent the E665 measurements. They conclude that the abundant production of hadrons in the target fragmentation region is due to cascade interactions. In the central rapidity region, additional production of hadrons due to multiple projectile collisions is seen in the $p \text{Xe}$ scattering, while little additional hadron production is seen in the $\mu \text{Xe}$ scattering \cite{Adams:1995}s. At large $n_{\text{g}}$, depletion of hadrons in the projectile fragmentation region is seen, presumably due to energy loss in projectile or cascade interactions. In the $\mu \text{Xe}$ scattering, the events with large $n_{\text{g}}$ account for only a tiny fraction of all events. Compared to the $p \text{Xe}$ data, the $\mu \text{Xe}$ data showed a stronger depletion of fast hadrons than would be expected from the number of projectile collisions.

In Fig.~\ref{fig:multiplicity-ratio-1} we can also see the multiplicity ratio simulated as a function of the number of grey tracks $n_{\text{g}}$ for $\hat{q}=0.5$ GeV$^{2}/$fm. The data consider up to  seven grey tracks, but in BeAGLE we predicted up to 12.

 \begin{figure*}[t!]
    \centering
    \includegraphics[width=\linewidth,angle=0]{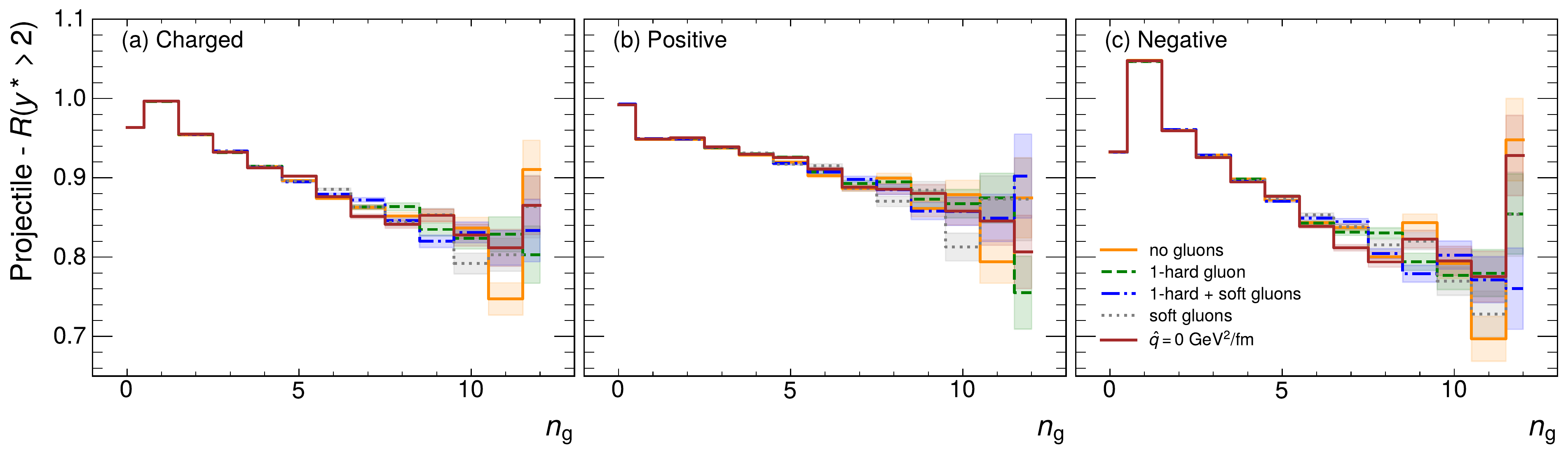}
    \caption{Multiplicity ratio $R(n_{g})_{\mu \text{Xe}}$ as a function of the number of grey tracks $n_{\text{g}}$ for charged, positive and negative hadrons for the projectile region considering the four PyQM options with transport coefficient $\hat q = 0.5$~GeV$^2$/fm. The result of simulations without induced energy loss with $\hat q = 0$~GeV$^2$/fm is also included as a reference. Charged, positive and negative particles are drawn in columns (a), (b) and (c), respectively. The different options of PyQM are highlighted by different types of lines and colors, where the orange (upper) solid line corresponds to the no gluons option, the green dashed-dotted line corresponds to 1 hard gluon option, the blue dashed-dotted line corresponds to 1 hard + soft gluons option, grey dotted line corresponds to soft gluons options, red (bottom) solid line corresponds to the simulation without induced energy loss with $\hat q = 0$~GeV$^2$/fm and full grey circles correspond to the E665 data.}
    \label{fig:multiplicity-ratio-2}
\end{figure*}


\begin{figure*}[t!]
    \centering
    \includegraphics[width=10cm,angle=0]{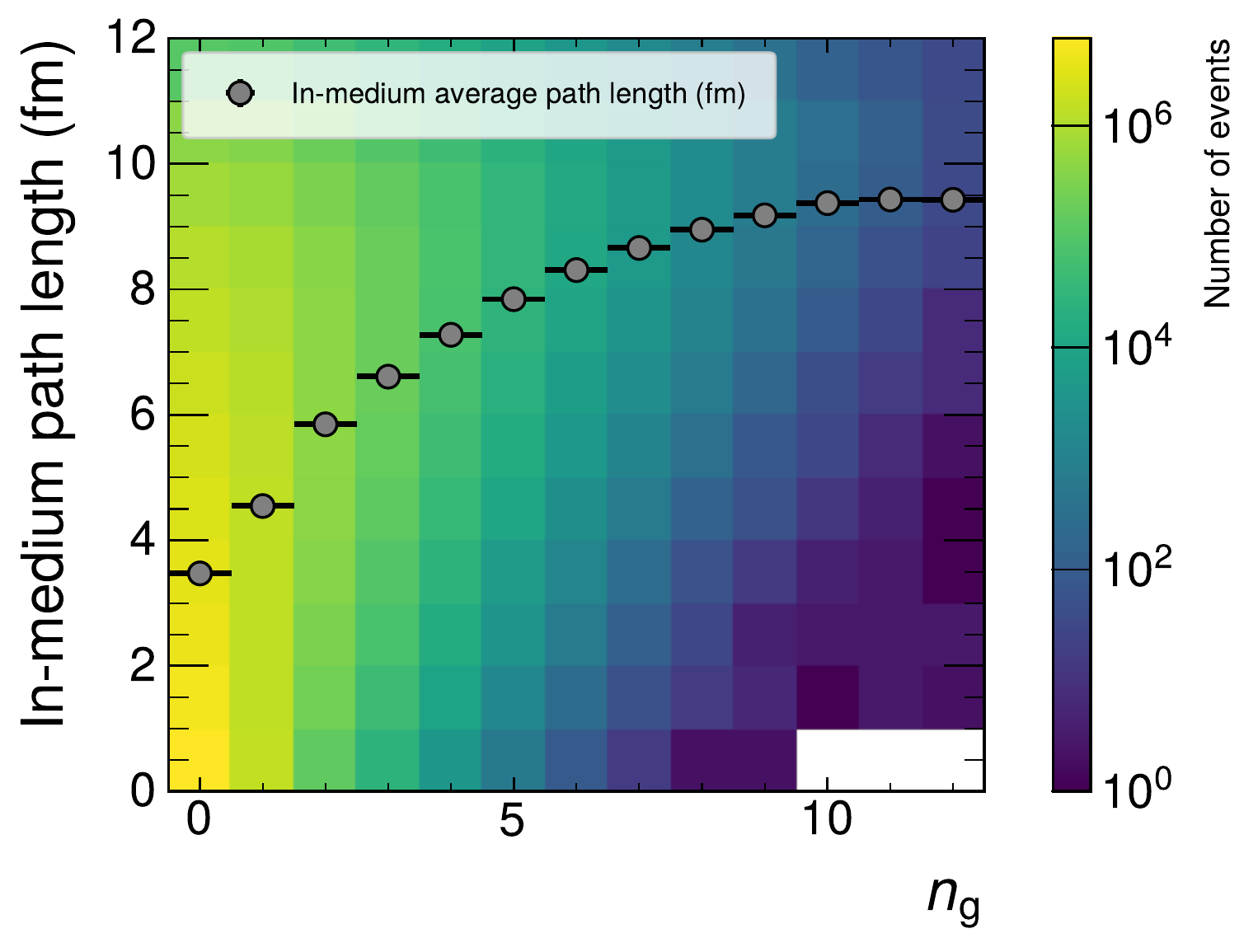}
    \caption{In-medium path length calculated by BeAGLE from the interaction point to the edge of the nucleus as a function of the number of grey tracks, $n_{\text{g}}$. This plot corresponds to BeAGLE with the option no gluons and transport coefficient $\hat q = 0.5$~GeV$^2$/fm.}
    \label{fig:distance}
\end{figure*}

The different colors indicate the different gluon radiation options in the PyQM module. Red represents no gluon radiation option, pink represents a hard gluon whose energy is $\Delta E=\omega_{\text{hard}}$, green represents the radiation of a hard gluon with at least energy $\omega_{\text{hard}}$ (eq. \ref{eq:whard}) plus soft gluons with energy $\omega_{\text{soft}}$ eq. (\ref{eq:wsoft}). The last option in blue indicates that only soft gluons are calculated, which means that all the energy is given to the remaining nuclei, taking into account that the energy of the soft gluons never exceeds 5~GeV.

To compare the different options and the effects of induced gluon radiation in a lepton-nucleus collision, we simulated different values of $\hat{q}$, but we only show $\hat{q}=0.5$~GeV$^{2}/$fm with $\hat{q}=0.0$~GeV$^{2}/$fm as reference.This is because even at this relatively large number, there is essentially no large effect due to energy loss. In Fig.~\ref{fig:multiplicity-ratio-1} we see the multiplicity ratios as a function of the grey tracks for each case and in Fig.~\ref{fig:multiplicity-ratio-2} we have rescaled the projectile region to see the small effects between the options and subsequently the energy loss.


In the target fragmentation region, BeAGLE underestimates the ratio for positive particles, where we also have a large number of grey tracks in this region, see Fig.~\ref{fig:multiplicity-ratio-1}. In the case of negative particles, BeAGLE predicts with high accuracy up to 5 grey tracks, but does not show much difference from any of the PyQM options. It shows very large ratios, but independent of energy loss, even with a high number of grey tracks, at least in this region.

In the projectile fragmentation region, BeAGLE accurately predicts up to four grey tracks for positive particles, but overestimates a high number of grey tracks without showing a large suppression effect, even for large values of $\hat{q}$. Therefore, our calculations of energy loss are not considerably different. For negative particles, BeAGLE continues to overestimate the ratios, with no sign of a large suppression for a small number of grey tracks.
 
The central fragmentation region provides a good description for negative particles, but not for positive particles, underestimating the ratios. In any case, this region is a combination of processes that also come from the other two regions.

In Fig.~\ref{fig:distance} we present the average number of grey tracks, and all grey tracks per event, compared to the in-medium path length of the struck quark. This length is calculated by BeAGLE from the interaction point to the edge of the nucleus. We observe a strong correlation between the grey tracks and the in-medium path length, despite the fact that the distribution of grey tracks is very large, indicating a general trend that more grey tracks are observed as the distance of the struck quark increases.

\FloatBarrier
\section{Conclusions}
 We find that the grey track production is dominated by interactions with in-medium hadrons in the backward region, where grey tracks are quite numerous. We have improved the PyQM module in BeAGLE to offer four options implementing partonic energy loss. Using a comparison of BeAGLE simulation to E665 grey track data we find that grey tracks are unaffected by such modifications for the forward production. We see a strong correlation between the number of grey tracks and the in-medium pathlength for lower values of $n_{\text{g}}$, an important quantity needed for precise modeling and interpretation of geometry-tagged data. This offers the advantage that a selection of certain particles in the forward region is unlikely to bias a centrality selection using the backward region grey tracks. 

Our energy loss model does not reproduce the suppression observed in the projectile region data, even with rather large values of $\hat{q}$. This raises questions on what could be at the origin of such a large suppression, unexpected at such high energies. We also see an unambiguous underprediction of the rate of positively charged grey track production in backward kinematics, suggesting that a much stronger interaction with the nuclear medium is needed. At the same time, there is very good agreement with the rate of negatively charged grey track production, indicating that the fragmentation process producing the negatively charged particles is well-described. 

These results lay an important foundation for future spectator tagging studies both with CLAS12 at Jefferson Lab, and at the Electron-Ion Collider, where both neutron and proton grey track studies will be feasible down to very small momenta. 

\section{Acknowledgments}
The authors thank Paul Hoyer and Stan Brodsky for stimulating and informative discussions concerning the physics processes of the parton-level scattering and production in the target reference frame and in the infinite momentum frame. The authors thank Elke Aschenaur for hosting a visit of CR to Brookhaven National Laboratory in the first stages of this work, and for support throughout the project. 
CR thanks Laboratoire de Physique des 2 Infinis Irène Joliot-Curie, CNRS - Université Paris-Saclay for extended research visits, and support throughout this project. This work was supported in part by the National Agency for Research and Development(ANID)/DOCTORADO BECA NACIONAL, Chile/2017-21171926, by PUCV DEA 1-2020, by UTFSM DGIIP 2-2020-2021, by ANID PIA grant ACT1413, by ANID PIA/APOYO AFB180002, by ANID FONDECYT No. 1161642, and by the ANID-Millennium Science Initiative Program - ICN2019\_044. 
This work was supported in part by the  U.S. Department of Energy contract DE-AC05-06OR23177, under which Jefferson Science Associates LLC manages and operates Jefferson Lab. AA also acknowledges support from DOE contract DE-SC0008791.
This work was partially supported by the Alexander von Humboldt Foundation. 
This work has also received funding from the European Research Council (ERC) under the European Union’s Horizon 2020 research and innovation programme (Grant agreement No. 804480).
Z. Tu is supported by the U.S. Department of Energy under Contract No. DE-SC0012704.
\FloatBarrier
\bibliography{Bibliography}
\end{document}